# EEG source localization using a sparsity prior based on Brodmann areas

S. Saha, Ya.I. Nesterets, Rajib Rana, M. Tahtali, Frank de Hoog and T.E. Gureyev

*Abstract*— Localizing the sources of electrical activity in the brain from Electroencephalographic (EEG) data is an important tool for non-invasive study of brain dynamics. Generally, the source localization process involves a high-dimensional inverse problem that has an infinite number of solutions and thus requires additional constraints to be considered to have a unique solution. In the context of EEG source localization, we propose a novel approach that is based on dividing the cerebral cortex of the brain into a finite number of "Functional Zones" which correspond to unitary functional areas in the brain. In this paper we investigate the use of Brodmann's areas as the Functional Zones. This approach allows us to apply a sparsity constraint to find a unique solution for the inverse EEG problem. Compared to previously published algorithms which use different sparsity constraints to solve this problem, the proposed method is potentially more consistent with the known sparsity profile of the human brain activity and thus may be able to ensure better localization. Numerical experiments are conducted on a realistic head model obtained from segmentation of MRI images of the head and includes four major compartments namely scalp, skull, cerebrospinal fluid (CSF) and brain with relative conductivity values. Three different electrode setups are tested in the numerical experiments.

## I. INTRODUCTION

UNDERSTANDING electrical activity inside human brain is potentially of great diagnostic value for epilepsy [1], stroke [2, 3], traumatic brain injury [4] and other brain disorders. Locating the sources of electrical activity inside the brain works by first finding the scalp potentials produced by virtual electric current dipoles at arbitrary locations in the brain (i.e. solving the forward problem), then, in conjunction with the actual EEG data measured by the electrodes, it is used to work back and estimate the sources that best fit the measurements (i.e. solving the corresponding inverse problem). In the cases where the number of measurement points (i.e. electrodes, usually <100 [5]) is lower than the number of unknowns (i.e. potential positions inside the head of the electrical dipoles with unknown current strength and orientation, >1000) this inverse problem is severely ill-posed [5] in the sense that there is an infinite number of source configurations that can produce the same distribution of the electric potential on the surface of the head. Hence additional constraints need to be introduced in order to find an appropriate unique solution. Note, however, that even with infinitely many data measurement points on the scalp, the spatial resolution of the EEG inversion will be limited due to the spreading of the electromagnetic signal on propagation through the head [6].

Various methods have been proposed for choosing suitable constraints for the inverse EEG problem, the most well-known being the 'minimum norm' (MN) constraint [7, 8]. Techniques relying on the MN constraint are based on a search for the solution with minimum power, along with regularization [5]. In other words, when the system is underdetermined, the solution is obtained by minimizing the $l_2$-norm of the solution components [9]. Several variants of this approach that consider different regularization parameters and weighting factors have already been proposed in the literature [5, 10]. Among them 'Standardized LOw REsolution brain Electromagnetic Tomography' (sLORETA) [11] has gained significant attention because of its capacity to ensure zero localization error at least in the case of a single source and noiseless environment. Despite its accuracy in the case of a single source and the relative simplicity of the corresponding computations which can produce the results very fast, sLORETA has been criticized for generating very broadly distributed or "smeared" sources in the reconstruction region [12] and for poor performance in the case of multiple simultaneously activated sources [13, 14].

In the last two decades, significant efforts have been made to develop new improved methods for solving ill-posed problems using sparse priors. FOCal Underdetermined System Solver (FOCUSS) is a classic example belonging to this category, which uses a weighted MN approach for sequentially reinforcing strong sources and suppressing the

S. Saha is a PhD student at the Department of Electrical Engineering in the University of New South Wales, Canberra, Australia and Intern in Biomedical Imaging at CSIRO Materials Science and Engineering, Melbourne, Australia (e-mail: S.Saha@student.adfa.edu.au, Sajib.Saha@csiro.au).
Ya.I. Nesterets is with CSIRO Materials Science and Engineering, Melbourne and with the University of New England, Australia (e-mail: Yakov.Nesterets@csiro.au).
Rajib Rana is with Autonomous Systems Laboratory, CSIRO ICT Centre, Australia (e-mail: Rajib.Rana@csiro.au).
M. Tahtali is with the School of Engineering and Information Technology, University of New South Wales, Canberra, Australia (e-mail: M.Tahtali@adfa.edu.au).
Frank de Hoog is with CSIRO Computational Informatics, Canberra, ACT, Australia (e-mail: Frank.Dehoog@csiro.au).
T.E. Gureyev is a Senior Principal Research Scientist in CSIRO, Australia and an Adjunct Professor at the University of New England, Australia (e-mail: Tim.Gureyev@csiro.au).

weak ones [15]. Other interesting algorithms are based on Iterative Reweighted Least-Squares (IRLS) methods, which are similar to FOCUSS and are based on iteratively computing weighted MN solutions with weights updated after each iteration [16, 17]. The homotopy method by Osborne [18] and LARS-LASSO algorithm [19, 20] (a variant of the homotopy method) are extremely powerful methods for solving the $l_1$ problem. Simple coordinate descent methods [21] or block wise coordinate descent, also called block coordinate relaxation [22], are also very successful strategies.

Following the discovery by Donoho and Candes *et al.* [23, 24] that sparsity could enable exact solution of ill-posed problems under certain conditions, there has been a tremendous growth of publications on efficient application of sparsity constraints for ill-posed problems [25, 26, 27, 28, 29, 45, 46, 47, 48]. Amongst these Zhang and Wu *et al.* specifically consider the EEG source localization problem. For example, Zhang has compared several state of the art sparse approximation algorithms for solving EEG source localization problem. Similarly, Wu *et al.* have proposed a matching pursuit based solution to the EEG inverse problem. While it produces better localization compared to the state-of-the art methods, the number of sources needs to be known a priori for the refinement of the localization in this method.

Despite the growing interest, the applicability of sparsity-enforcing priors for EEG source localization is still limited because of the significantly smaller number of electrodes compared to the typical number of virtual electric current dipoles in consideration. Wu *et al.* algorithm is the best to our knowledge and have only been able to accurately locate up to 6-8 active dipoles based on a 61-electrodes setup and considering only 1279 dipoles [26]. For a more realistic head model, which commonly uses about 6000 dipoles, with each dipole corresponding to about 5×5×5 mm³ of grey matter, this method is expected to give worse results for a similar electrode setup, because of the increased number of unknowns. Importantly, it is well accepted in the literature that a region of the brain corresponding to a group of virtual dipoles, rather than to a single dipole, tends to be activated during a certain brain activity [6]. For example, it is well known that Brodmann area 17 is related to human visual activity. In a model used in [14] with 6203 dipoles, the number of dipoles that belong to that region is about 54. Therefore, even the best sparsity-based reconstruction proposed so far will likely fail to locate the activity accurately, if a group of more than 8 dipoles in that area are activated simultaneously. It has been pointed out by Wagner and colleagues [13] that in the sLORETA method, closely located activated dipoles will produce a broad region having activity maxima located somewhere in between, therefore sLORETA may produce acceptable results in such a scenario. At the same time, sLORETA will likely produce inaccurate results when more than one group of dipoles corresponding to different spatially separated areas of the brain are activated simultaneously. Assuming typical activation of the group of dipoles rather than a single dipole and considering the severely underdetermined nature of the problem the sparsity-based reconstruction methodology proposed in this paper groups the dipoles based on their activity and then applies sparsity constraint for detection of activated groups or region(s) in the brain. Since the full functionality of different parts of the human brain is still not known completely, the grouping of dipoles here is based on Broadmann areas - a cytoarchitectural, rather than a functional, classification of the human cerebral cortex. However, even though Brodmann areas were originally defined on the basis of the cytoarchitectural organization of neurons in the cerebral cortex, more recent studies have unveiled the structural–functional correlations of many Broadmann areas and thus point to the potential suitability of such segmentation of the brain as a tentative basis for a functional classification relevant to EEG [30].

Three different EEG headset configurations with 19, 33 and 71 electrodes are used for the numerical experiments in this paper. A schematic representation of the headsets in consideration is shown in Figure 1.

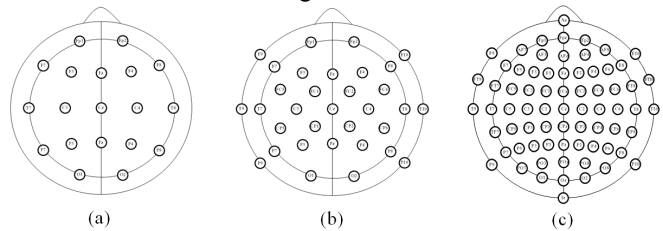

Fig. 1. Schematic representation of the electrodes positions in the (a) 19-Electrodes (b) 33-Electrodes (c) 71-Electrodes EEG headset configurations.

## II. BACKGROUND

### A. Mathematical Formulation of EEG imaging

The equation often used for defining the forward and inverse imaging problems in EEG has the following form:
$$\boldsymbol{\Phi} = \boldsymbol{K}\boldsymbol{J} + c\mathbf{1}. \qquad (1)$$
Here $\boldsymbol{\Phi} \in \mathbb{R}^{N_E \times 1}$ is a vector of the scalp electric potentials measured by the $N_E$ electrodes with respect to a reference electrode, $\boldsymbol{J} \in \mathbb{R}^{3N_V \times 1}$ is the primary or impressed current density vector, where $N_V$ is the number of dipole locations in the brain, with each dipole current vector having three independent components corresponding to the usual Cartesian coordinates in 3D space, $\boldsymbol{K} \in \mathbb{R}^{N_E \times 3N_V}$ is the lead field matrix, c is a constant which embodies the fact that the electric potential is determined up to an arbitrary constant [11] and $\mathbf{1} \in \mathbb{R}^{N_E \times 1}$ is a vector of ones.

Source localization works by first calculating the scalp potentials produced by virtual dipoles at arbitrary locations in the brain (i.e. solving the forward problem), and then, in conjunction with the actual EEG data measured by the electrodes, it is used to work back and estimate the sources that best fit the measurements (i.e. solving the corresponding inverse problem). Typically, this inverse problem is ill-posed, hence additional constraints are added in order to produce an appropriate unique solution.

### B. Compressed Sensing and Sparse Signal Recovery

Compressed sensing is an emerging field of information theory which shows that one can exploit sparsity[1] or compressibility when acquiring signals of general interest and that one can design a non-adaptive sampling strategy that condenses the information in a compressible signal into a small amount of measurements. In a nutshell, compressed sensing proposes to find a signal $\boldsymbol{J} \in \mathbb{R}^N$ by collecting $M$ linear measurements of the form $\Phi_m = <\boldsymbol{K}_m, \boldsymbol{J}> + z_m, 1 \leq m \leq M$ or in matrix notation

$$\boldsymbol{\Phi} = \boldsymbol{K}\boldsymbol{J} + \boldsymbol{z} \quad (2)$$

where $\boldsymbol{K}$ is a $N \times M$ sensing matrix with $N$ usually smaller than $M$ (by one or several orders of magnitude, i.e., $N<<M$) and $z$ is an error term representing measurement error. Since $N < M$, eq. (2) is an underdetermined linear system and such systems do not have a unique solution. Encouragingly, sparsity constraint is useful to find a unique solution of eq. (2) [31]. In particular, one can solve the following optimization problem to find a unique solution of eq. (2)

$$\widehat{\boldsymbol{J}_0} = \arg \min \| \boldsymbol{J} \|_0 \text{ s.t. } \boldsymbol{\Phi} = \boldsymbol{K}\boldsymbol{J}. \quad (3)$$

Here $\| \boldsymbol{J} \|_0$ refers to the $l_0$-norm, which counts the number of nonzero elements in a vector. If a signal has less than $N/2$ non-zero elements it is possible to find a unique sparsest solution [31]. However, finding the $l_0$-norm of an underdetermined system is NP hard[2] [32]. Encouragingly, it has been shown in the compressive sensing literature that if the solution is sparse enough, it is possible to estimate the $l_0$-norm by solving the following $l_1$-norm optimization problem:

$$\widehat{\boldsymbol{J}_1} = \arg \min \| \boldsymbol{J} \|_1 \text{ s.t. } \boldsymbol{\Phi} = \boldsymbol{K}\boldsymbol{J}, \quad (4)$$

when $N \geq S \log (M/S)$.

Another important consideration for sparse approximation is the so-called coherence of the projection matrix $\boldsymbol{K}$ [33] which is defined as

$$\mu(\boldsymbol{K}) = \max_{i<j} \frac{|<K_i, K_j>|}{\|K_i\|_2 \|K_j\|_2}, \quad (5)$$

where $K_i$ and $K_j$ denote the $i$-th and $j$-th column of $\boldsymbol{K}$, respectively. We say that a matrix is incoherent if $\mu$ is much smaller than one. Generally lower coherence ensures better reconstruction from a small number of measurements.

Sparse signal recovery algorithms can be categorized into greedy algorithms [34], algorithms based on mixed norm optimization [35], iterative reweighted algorithms [36] and Bayesian algorithms [37]. Sparse Bayesian Learning (SBL) methodology [wirf04] has been found very promising for solving the inverse problems when the projection matrix is highly coherent [25], which is typical for EEG. Considering the highly coherent nature of the lead-field matrix in this work we have used SBL [38] to solve eq. (4) in conjunction with the Broadmann map based segmentation of the cerebral cortex. A number of algorithms based on SBL also consider incorporating temporal correlation of the sources [39] to ensure better performance, however in this work we do not consider any temporal correlations because of the additional complexity of computing and then efficiently incorporating such information, which could be a computationally demanding task [39].

*C. Brodmann Areas*

Brodmann areas were originally defined by the German anatomist Korbinian Brodmann in 1909 based on the cytoarchitectural organization of neurons that he observed in the cerebral cortex as a result of a careful microscopic examination [40]. Brodmann's map of the human cortex contains 43 cytoarchitectonic areas, which are labeled by numbers between 1 and 52. Areas with the numbers 12–16 and 48–51 are not used in his map of the human brain [30]. Brodmann explained these 'gaps' by the fact that some areas are not identifiable in the human cortex but are well developed in other mammalian species. Brodmann areas have been discussed, debated, refined, and renamed exhaustively for more than a century. In the 1980s Brodmann's map gained a renewed popularity with the introduction of novel functional and structural neuroimaging techniques, which allowed the translation of the two-dimensional information of the original map into a three-dimensional representation [30]. The goal of the Brodmann's map was to produce a comparative organic theory of the cerebral cortex based on anatomical features [30]. This hypothesis could not be rigorously tested in his time, except for some so-called primary areas such as the primary visual cortex. Recent functional imaging studies have demonstrated that in many cases this anatomical segmentation of the human brain is valid for its functionality as well, although there are studies demonstrating that the map could be incomplete in other cases [30].

It is already well accepted that neurons in the brain work in groups or clusters, however no precise map is currently available that would be able to group the neurons accurately in regard to brain functionality. We therefore rely in the present work on the most widely known and accepted cytoarchitectural segmentation of the human cortex represented by the Brodmann's map.

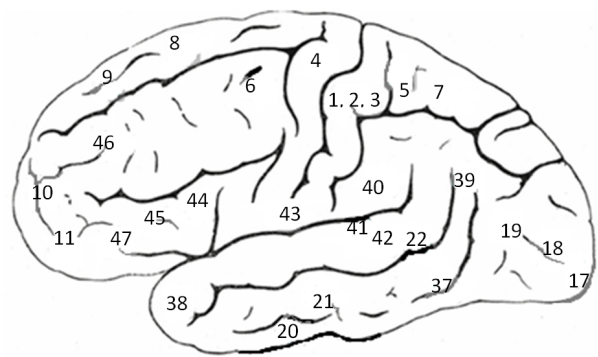

Fig. 2. Lateral Surface of the Brodmann map.

---

[1] A signal $x$ is called $S$-sparse if it has only $S$ nonzero elements. However, natural signals are rather compressible. A signal $x$ is designated compressible if it has only a small proportion of large coefficients when the signal is transformed into a suitable domain such as, Haar, Fourier etc. Mathematically, a signal is compressible if the coefficients decay obeys the power law [candes05]

[2] NP-hard means at least as hard as any 'NP-problem' (nondeterministic polynomial time problem); it might, in fact, be harder.

## III. Proposed Approach

In order to better localize the sources of electrical activity inside human brain the proposed model relies on the following two steps

- Firstly it considers grouping the dipoles into several zones based on functionality. Such consideration reduces the search space significantly and thus points to more accurate activity localization.

- Secondly it applies TMSBL [39] algorithm to reconstruct the sparse signal. It relies on the hypothesis that the signal is sparse when it is presented in terms of the active Functional Zones.

### A. Data Model and Assumptions

A realistic head model suitable for solution of the direct and inverse EEG problems include four different major components, namely scalp, skull, CSF and brain, with the following relative conductivity values [6]: $\sigma_{scalp}=1$, $\sigma_{skull}=0.05$, $\sigma_{CSF}=5$, $\sigma_{brain}=1$. The source space was constructed by dividing the head model into cubes with a size of $5\times5\times5$ mm$^3$ and considering possible current dipoles only in the center of those cubes that consisted of at least 60% of gray matter. This segmentation procedure resulted in 6203 dipole positions. In order to implement the proposed sparsity criteria, the considered 6203 dipoles were clustered according to Brodmann areas. While the number of Brodmann areas for the human cortex is 43 [30], the above mentioned segmentation procedure produced no dipoles in BA 26 and thus resulted in 42 Brodmann areas in total. Considering both the left and the right hemisphere, we ended up with 84 dipole clusters which we called Functional Zones. We assumed that all the dipoles (or most of the dipoles) in a particular Functional Zones are activated simultaneously during specific tasks performed by the human brain. The considered 84 Functional Zones were used to recompute the previously calculated lead field matrix from $K \in \mathbb{R}^{N_E \times (3 \times 6203)}$ to $K_{FZ} \in \mathbb{R}^{N_E \times (3 \times 84)}$, by taking the average of all the lead field values belonging to each group in regard to $x$, $y$ and $z$ components of each electric current dipole.

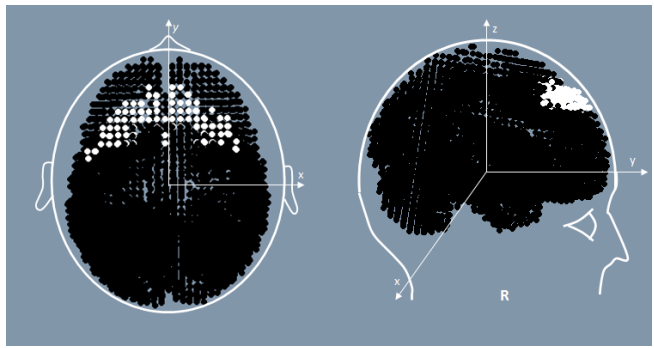

Fig. 3. Brodmann area 8 (white dotted) shown over the original subdivision of the cortex into 6203 dipoles.

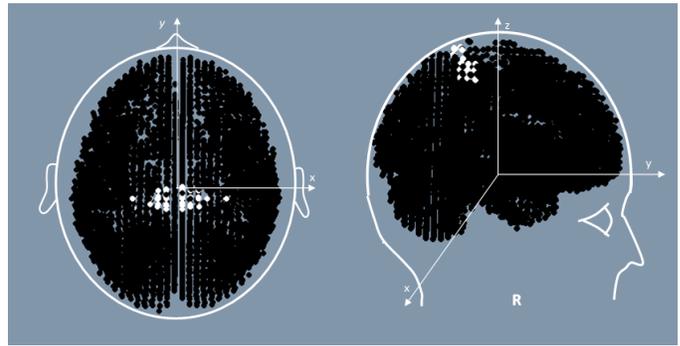

Fig. 4. Brodmann area 5 (white dotted) shown over the original subdivision of the cortex into 6203 dipoles.

### B. Localizing the Sources of Electrical Activity

Sparse Bayesian Learning (SBL) methodology [38] is used to find the inverse solution. SBL was initially proposed for regression and classification by Tipping [37] in machine learning. In [38], Wipf et al. applied SBL for the sparse signal recovery problem. The idea of SBL is to find $J$ through Maximum a Posterior (MAP) estimate [39]. Several variants of SBL algorithm are found in the literature [41], and in this work we consider the TMSBL [39] algorithm, which is found superior when the dictionary matrix is highly coherent [25]. While TMSBL is derived for Multiple Measurement Vector (MMV) model, it can be used in the Single Measurement Vector (SMV) model (i.e. our case). In this case TMSBL [39] is similar to EM-SBL [38] with the key difference in the error variance ($\sigma^2$) learning rule [41]. The parametric form of the SBL weight prior here is given by

$$p(\boldsymbol{J},\boldsymbol{\gamma}) = \prod_{i=1}^{M} (2\pi\gamma_i)^{-\frac{1}{2}} \exp(-\frac{J_i^2}{2\gamma_i}),$$

where $\boldsymbol{\gamma} = [\gamma_1,...,\gamma_M]^T$ is a vector of $M$ hyperparameters controlling the prior variance of $J$. These hyperparameters along with the error variance ($\sigma^2$) are estimated from the data by marginalizing over $J$ and then performing maximum likelihood optimization [38]. The marginalized probability density function is given by

$$p(\boldsymbol{\Phi},\boldsymbol{\gamma},\sigma^2) = \int p(\boldsymbol{\Phi}\mid\boldsymbol{J};\sigma^2) p(\boldsymbol{J};\boldsymbol{\gamma}) d\boldsymbol{J}$$

$$= (2\pi)^{-\frac{N}{2}} \mid\Sigma_\Phi\mid^{-\frac{1}{2}} \exp[-\frac{1}{2}\boldsymbol{\Phi}^T \Sigma_\Phi^{-1} \boldsymbol{\Phi}],$$

where $\Sigma_\Phi = \sigma^2 \boldsymbol{I} + \boldsymbol{K}\boldsymbol{\Gamma}\boldsymbol{K}^T$, $\boldsymbol{\Gamma} = \begin{bmatrix} \gamma_1 & & & \\ & \cdot & & \\ & & \cdot & \\ & & & \cdot \\ & & & & \gamma_M \end{bmatrix}$.

We now need to estimate $\gamma$ and $\sigma^2$ from the data and for this we employ EM algorithm [39]. Once we have these values the solution of the inverse problem is computed as

$$\hat{\boldsymbol{J}} = (\boldsymbol{K}^T\boldsymbol{K} + \sigma_{ML}^2 \boldsymbol{\Gamma}_{ML}^{-1})^{-1} \boldsymbol{K}^T \boldsymbol{\Phi}$$

The solution $\hat{J}$ is a (3×84)×1 vector represents current sources at the 84 locations within the brain volume with three directional (i.e. *x, y, z* directions) components per location. The *x*, *y* and *z* components are used to calculate the magnitude of the current density for each of the Functional Zones. From the magnitudes of the current density of the Functional Zones, the maximum magnitude, *Max_Mag* is determined. Any Functional Zone with a magnitude larger than or equal to *thr* of *Max_Mag*, is considered to be active. The threshold, *thr* is experimentally set to 20% of *Max_Mag*, and provides a good choice to suppress unwanted maxima (if there is any) keeping the actual maxima.

IV. EXPERIMENTAL ANALYSIS

Experiments were conducted by varying the number of simultaneously active Functional Areas for three EEG headset configurations as specified in the 'Introduction' section. The idea of using different electrodes setups for the experiment was to analyse how the reconstruction performance of the proposed approach varies from low density to high density headsets.

Error Distance (ED) [42] was used to analyse the reconstruction performance. The error distance between the actual and the estimated source locations is defined as

$$ED = \frac{1}{N_I} \times \sum_{i \in I}^{N_I} \min_l ||r_i - s_l|| + \frac{1}{N_L} \times \sum_{l \in L}^{N_L} \min_i ||s_l - r_i|| \quad (6)$$

Here $s_l$ and $r_i$ are the actual and estimated source locations respectively. $N_I$ and $N_L$ are the total numbers of estimated and the undetected sources respectively. The first term of equation (6) calculates the mean of the distance from each estimated source to its closest real source, and the corresponding real source is then marked as detected. All the undetected real sources made up the elements of the data set *L* and thus the second term of the equation calculates the mean of the distance from each of the undetected sources to the closest estimated source.

*A. Localization Error - Single Source Activation*

First, a single Functional Zone was considered active and the corresponding potentials on the electrodes were calculated based on $K_{FZ}$. The experiment was conducted for all Functional Zones activated sequentially one at a time. Localization of the Functional Zone was performed according to the algorithm specified above in section 'Localizing the Sources of Electrical Activity'. For this experiment, each Functional Zone was represented by its centroid $s_i (\forall_i \in 1:FZ)$. For an active source position $s_i$ we claimed a "success" if the error distance was zero. The success rate was computed as $\frac{\sum_{i=1}^{AZ}(ED^i==0)}{FZ}$. We computed mean error distance, $E = \sum_{i=1}^{FZ} ED^i / N_{unsuccessful}$ for unsuccessful cases as localization error. Since we assumed that each Functional Zone could have random "direction of activity" (i.e. the orientation of the average dipole moment), the whole experiment was conducted 100 times. For each run we computed the success rate, the mean error distance (considering only unsuccessful cases) and the standard deviation. Table I shows the average of 100 such findings.

TABLE I
SUCCESS RATE AND THE LOCALIZATION ERROR ANALYSIS FOR UNSUCCESSFUL CASES FOR ONE ACTIVE ZONE

| Number of Electrodes | Success rate | Mean error distance (for unsuccessful cases) (mm) | Standard deviation (for unsuccessful cases) (mm) |
|---|---|---|---|
| 19 | 0.86 | 20.62 | 11.96 |
| 33 | 0.98 | 14.94 | 10.27 |
| 71 | 1.00 | 0 | 0 |

*B. Localization Error - Multiple Sources Activation*

In this case *S* (*S* >1) Functional Zones were activated simultaneously. From the total of $C_S^{84} = \frac{84!}{S!(84-S)!} = \frac{84(84-1)(84-2)...(84-S+1)}{S!}$ possible combinations of 84 Functional Zones, one combination was chosen randomly. For each value of *S* we claimed a "success" if the error distance was zero. Since Functional Zones could have random directions of activity, the experiment was performed 1000 times and the success rate was computed over these 1000 runs for each value of *S*. For the unsuccessful cases we computed the mean error distance, $E = \frac{\sum_{i=1}^{1000} ED^i}{N_{unsuccessful}}$ and the standard deviation of the mean error distance for each value of *S*. Figure 5 shows the success rates as a function of *S*. Table II summarizes the localization errors for unsuccessful cases.

From the results it is observable that the success rate decreases with the increased number of simultaneously active zones. Unsurprisingly, better localization accuracy was found for the 71-electrodes setup in comparison to other setups. Even for 6 simultaneously active Functional Zones, the 71-electrodes setup was able to locate all the active zones accurately in around 70% cases, for the rest 30% cases it produced localization error of about 15 mm. For 33-electrodes setup and up to 3 simultaneously active Functional Zones the proposed approach was found to locate the active zones accurately for >=70% cases (with a localization error of <=15 mm for unsuccessful cases). Of course, larger numbers of simultaneously active Functional Zones result in less accurate localizations. For 19-electrodes setup the proposed approach produced accurate localization only for 40% cases even only for 2 simultaneously active Functional Zones with a localization error of about 21 mm (for the rest 60% cases).

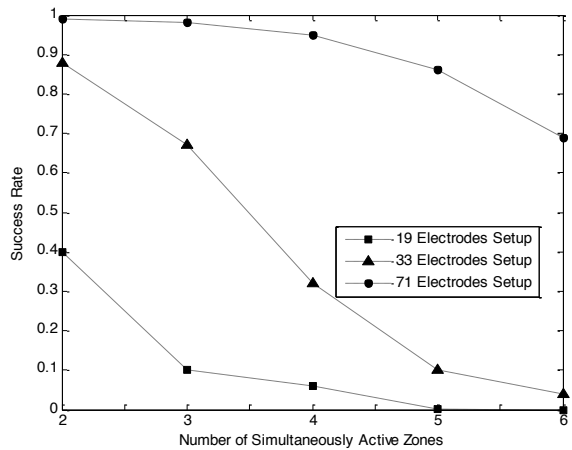

Fig. 5. Influence of the number of simultaneously active areas on the performance of the reconstruction.

TABLE II
LOCALIZATION ERROR ANALYSIS FOR UNSUCCESSFUL CASES, WHERE "SUCCESS" MEANT THAT ALL THE ACTIVATED FUNCTIONAL ZONES WERE EXACTLY LOCATED IN THE RECONSTRUCTED SIGNAL

| Number of simultaneously active areas | 19 Electrodes Setup | | 33 Electrodes Setup | | 71 Electrodes Setup | |
|---|---|---|---|---|---|---|
| | Mean error distance (mm) | Std. (mm) | Mean error distance (mm) | Std. (mm) | Mean error distance (mm) | Std. (mm) |
| 2 | 20.80 | 11.05 | 14.11 | 9.91 | 5.16 | 4.64 |
| 3 | 26.19 | 20.33 | 15.22 | 10.24 | 4.97 | 4.35 |
| 4 | 39.72 | 22.41 | 16.78 | 15.18 | 8.03 | 7.43 |
| 5 | 51.27 | 23.15 | 27.29 | 20.83 | 11.04 | 9.84 |
| 6 | 57.47 | 21.98 | 34.23 | 20.25 | 16.53 | 15.19 |

### C. Partially activated Functional Zone

This part of the experiment was designed to analyse the performance of the localization method when only a fraction of the Functional Zone rather than the whole Functional Zone was active. For each of the activated Functional Zone a specified percentage of the dipoles belonging to that Functional Zone were activated. One dipole of the considered Functional Zone was chosen randomly and the rest of the active dipoles (based on the specified percentage of activated dipoles) were adjacent to the chosen dipole. The Lead field matrix $K$ was used for the forward problem (i.e. to generate $\Phi$), whereas $K_{FZ}$ was used for the inverse problem. While Functional Zone(s) can have random direction of activity, all the considered dipoles within a given Functional Zone were considered to have the same orientation. The experiment was conducted for all the Functional Zones ($1:FZ$) activated sequentially, one at a time, for the 33 electrode setup. In this case we claimed a "success" if the activated Functional Zone was exactly detected in the reconstructed signal (i.e. if the error distance was zero). The success rate was computed as $\frac{\sum_{i=1}^{FZ}(ED^i==0)}{FZ}$ and mean error distance, $E$ was computed as $\sum_{i=1}^{FZ} ED^i / N_{\text{unsuccessful}}$ considering only unsuccessful cases. The whole experiment was conducted 10 times and for each run we computed the success rate and the mean error distance (considering only unsuccessful cases). The results shown in Table III are the average over 10 such findings.

TABLE III
SUCCESS RATE AS A FUNCTION OF THE ACTIVATED PERCENTAGE OF A FUNCTIONAL ZONE (FOR 33 ELECTRODES SETUP)

| Percentage of the active area | Success rate | Mean error distance (for unsuccessful cases) (mm) |
|---|---|---|
| 100 % | 0.99 | 14.83 |
| 90 % | 0.94 | 15.05 |
| 80 % | 0.73 | 14.68 |
| 70 % | 0.54 | 15.69 |
| 60 % | 0.47 | 17.05 |
| 50 % | 0.38 | 18.03 |

From the results it is clear that when the whole area of the considered zone is active it is very likely that it will be exactly localized in the reconstructed signal. As soon as the percentage of the active area decreases the chances of inexact localization increases. Another important observation is that when the percentage of activation decreases one could have more than one active Functional Zone in the reconstructed signal, however it is very likely that the activity maxima will still point to the actual activated Functional Zone. In order to verify that claim, rather than computing the error distance as in Table III, we considered only one Functional Zone having the maximum magnitude in the reconstructed signal and then computed the Euclidian distance between the actual and estimated Functional Zone. The findings are shown in Table IV.

TABLE IV
SUCCESS RATE IN REGARD TO PERCENTAGE OF THE ACTIVATE AREA (FOR 33 ELECTRODES SETUP). THE RESULTS SHOWN HERE IS THE AVERAGE OVER 10 RUNS.

| Percentage of the active area | Success rate | Mean Euclidian distance (for unsuccessful cases) (mm) |
|---|---|---|
| 100 % | 0.99 | 16.05 |
| 90 % | 0.98 | 17.03 |
| 80 % | 0.95 | 16.87 |
| 70 % | 0.89 | 17.56 |
| 60 % | 0.84 | 17.89 |
| 50 % | 0.78 | 21.34 |

### D. Noise Sensitivity

Here we investigated the effect of simulated pseudo-random measurement noise superimposed on the measured scalp potentials. First, a single Functional Zone was considered active and the corresponding potentials on the electrodes were calculated based on $K_{FZ}$. We then added variable amounts of noise to each of the electrode potentials. This random noise was assumed to be statistically independent Gaussian noise with zero mean and standard deviation $\sigma$. For a given Signal to Noise Ratio (SNR), the standard deviation was computed as $\sigma = \max\{(P_{Signal}^e)\} / \text{SNR}$ where $P_{Signal}^e$ is the signal measured by the electrode $e$ ($\forall_e \in 1:N_E$). The experiment was conducted for all Functional Zones activated sequentially one at a time and for different levels of SNR (5, 10, 15, 20, 25, 30 and infinity). Localization of the Functional Zone was performed according to the algorithm specified above in section 'Localizing the Sources of Electrical Activity'. As we were adding a random amount of noise, the above mentioned experiment was conducted 100 times for each Functional Zone, $s_i(\forall_i \in 1:FZ)$. Figure 6 shows the success rate of the experiment for different electrodes setups and different levels of SNR.

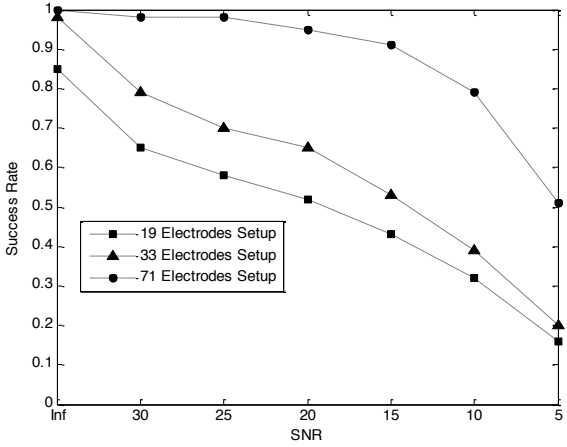

Fig. 6. Success rate for different levels of SNR, in the case of single Functional Zone.

From the results it is observable that the success rate decreases with the increase in the amount of noise. Unsurprisingly, the 71-electrodes setup is the most robust against noise in localizing the sources of activity accurately. For the 71-electrodes setup the proposed approach was found > 50% successful even with SNR of 5. For 31- and 19-electrodes setup the proposed method produced success rate of greater than and around 50% for the SNR of 15.

*E. Analysis of the lead-field matrices (Conditioning of the reconstruction)*

Note that equation (1) is ill conditioned in two respects. Firstly, it is under determined in the sense that there are fewer equations than unknowns, although, to a certain extent, this is ameliorated here by the assumption that the solution is sparse. Secondly the columns of the lead field matrix, $K$ are highly coherent, which suggests that the error in the data will be amplified when solving for $J$.

For $l_2$ approximation, such ill conditioning can be analysed via the singular value decomposition of the lead field matrix $K$ [43], namely $K = USV^T$ where $S = \mathrm{diag}(s_1, \ldots, s_{N_E})$, $s_1 \geq \cdots \geq s_{N_E} \geq 0$ and $U \in \mathbb{R}^{N_E \times N_E}$, $V \in \mathbb{R}^{N_E \times 3N_V}$ satisfy $UU^T = I$ and $VV^T = I$. The singular values for the lead field matrix based on the Brodmann areas are given in Figure 7. Although we are interested in $l_0$ approximation rather than $l_2$ approximation, it is plausible that the singular values should play a role here also.

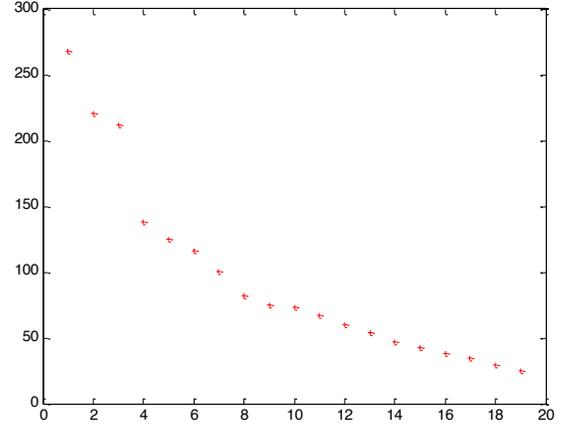

(a)

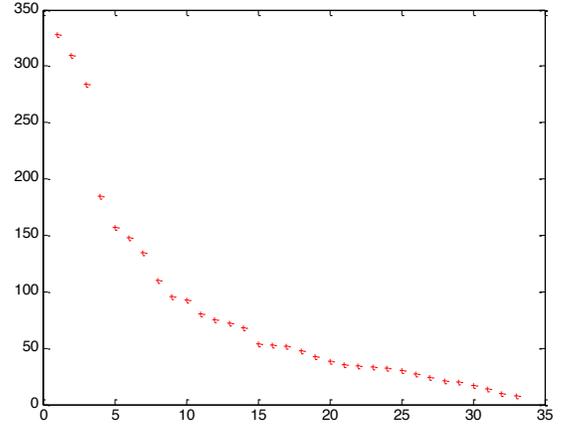

(b)

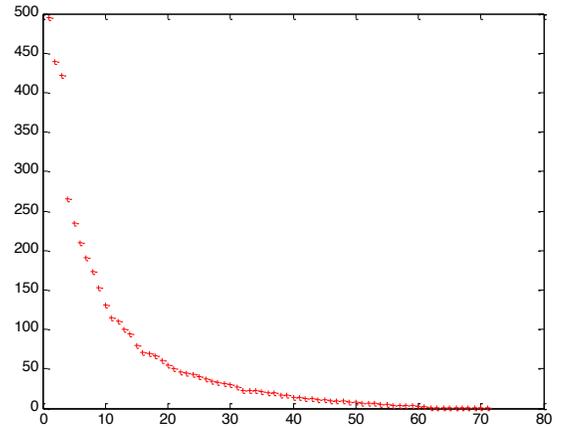

(c)

Fig. 7. Singular values of lead field matrix for Brodmann areas with, (a) 19 electrodes, (b) 33 electrodes, (c) 71 electrodes.

To investigate this, we note that equation (1) can be rewritten as

$$V^T J = S^{-1} U^T (\Phi - c\mathbf{1}) \qquad (6)$$

and we compare the errors in (1) to the effect of errors in (6). Specifically, we compare the effect of the perturbed equation

$$KJ = \Phi + \alpha^{-1}||\Phi||_2 \varepsilon, \qquad (7)$$

or, equivalently,

$$\boldsymbol{V}^T \boldsymbol{J} = \boldsymbol{S}^{-1} \boldsymbol{U}^T \boldsymbol{\Phi} + \beta^{-1} ||\boldsymbol{S}^{-1} \boldsymbol{U}^T \boldsymbol{\Phi}||_2 \boldsymbol{\varepsilon}. \qquad (8)$$

Here $\boldsymbol{\Phi}$ is the measured potential when only one of the Brodmann areas is active, the components of $\boldsymbol{\varepsilon}$ of are mean zero independently and identically distributed random numbers with variance $1/N_E$, $\alpha$ is the signal to noise ratio for equation (7) and $\beta$ is the signal to noise ratio for equation (8).

We have performed a number of simulations aimed at quantifying the accuracy of identifying which Brodmann area is active when only one area is active and we have found that the accuracy for (7) with the signal to noise ratio $\alpha = 10$ when the number of electrodes is 33 gives comparable results to (8) when $\beta = 2.2$. A simple analysis indicates that, for the general case, the performance of (7) and (8) are the same when $\beta = \kappa \alpha$, where

$$\kappa = N_E \left( \sum_{i=1}^{N_E} s_i^{\,2} \right)^{\frac{1}{2}} \left( \sum_{i=1}^{N_E} s_i^{\,-2} \right)^{\frac{1}{2}}$$

Thus, the term $\kappa$ provides a measure of the effect of conditioning of the lead field matrix.

## V. Conclusions

Recent years have witnessed a tremendous interest to sparse approximation in signal processing. Sparsity constraints have also been considered in EEG for localizing the sources of electrical activity inside the brain, i.e for solving the inverse EEG problem. Considering that the number of dipoles in a realistic head model is usually significantly higher than the typical number of electrodes in an EEG headset, and the electrical signals spread very significantly on propagation through the head, this inverse problem is highly ill-posed, and the best algorithms are able to solve the problem only for a very limited number of simultaneously active dipoles. It is already well accepted that usually a group of adjacent dipoles rather than a single dipole gets activated simultaneously. Hence in order to better specify the sparsity profile, our proposed method groups the considered dipoles in the human cerebral cortex into several Functional Zones, each zone corresponding to one of the Brodmann's areas, and then applies the sparsity constraint to these Functional Zones rather than individual dipoles.

The proposed method was tested by varying the number of simultaneously activated Functional Zones and with different electrode setups. The results indicate that the proposed method is quite promising in locating the activated Functional Zones accurately. For a 71-electrodes setup and up to 2 simultaneously activated Functional Zones, the proposed model can locate the sources of activation exactly in the noiseless case. Even though the percentage of exact localization decreases with the increase of the number of simultaneously active zones, the percentage of exact localization is still around 70% up to 6 simultaneously activated Functional Zones. Unsurprisingly, for the 31- and the 19-electrodes setups the localization accuracy is not as high as for the 71-electrodes setup; however when the signal is highly sparse (e.g. up to 4 simultaneously activated Functional Zones for 31-electrodes setup, and up to 2 simultaneously activated Functional Zones for 19-electrodes setup), these setups are found to produce fairly accurate localization. Since the relevant details of the human brain activity are still not known precisely, it is particularly important to know what percentage of the area of the Functional Zone needs to be activated for successful localization. Experiments reveal that the proposed model is fairly accurate to locate the sources of activity even with 50% activation of the Functional Zone.

While the proposed model relies on Brodmann's map to define the Functional Zones, it is will be interesting to consider other widely used brain atlases, for example AAL [44] which may better correlate with a typical functional segmentation of the human cortex. Our future work will aim at analysing the performance of the proposed approach in regard to different brain atlases and evaluating the correlation of the EEG-based activity localization with that of the fMRI.

ACKNOWLEDGMENT

This project was supported by the Computational and Simulation Sciences Transformational Capability Platform of Commonwealth Scientific and Industrial Research Organisation (CSIRO), Australia, along with University of New South Wales (UNSW), Canberra, Australia. The authors would like to thank Dr. Chao Suo and Dr. Roger Koenig-Robert of Monash Biomedical Imaging (MBI) Laboratory, VIC, Australia for their valuable comments and suggestions.